\documentclass{aadebug}
\usepackage[psamsfonts]{amssymb}
\usepackage{graphicx}
\usepackage{type1cm}
\usepackage{times, mathptmx}

\newcommand{\lw}[1]{\smash{\lower1.5ex\hbox{#1}}}

\newcommand{\sor}{\stackrel{S}{\rightarrow}}
\newcommand{\cor}{\stackrel{C}{\rightarrow}}
\newcommand{\nor}{\stackrel{N}{\rightarrow}}
\newcommand{\hbr}{\rightarrow}

\begin{document}
\corr{0309027}{127}

\runningheads{Masao Okita et al.}{Debugging Tool for Localizing Faulty Processes in Message Passing Programs}

\title{Debugging Tool for Localizing Faulty Processes in Message Passing Programs}

\author{
Masao~Okita\addressnum{1}\comma\extranum{1},
Fumihiko~Ino\addressnum{1}\comma\extranum{1},\\
and Kenichi Hagihara\addressnum{1}\comma\extranum{1}
}

\address{1}{
Graduate School of Information Science and Technology,
Osaka University\\
1-3 Machikaneyama,
Toyonaka,
Osaka 560-8531,
Japan
}
\extra{1}{E-mail: \{m-okita,ino,hagihara\}@ist.osaka-u.ac.jp}

\pdfinfo{
/Title (Debugging Tool for Localizing Faulty Processes in Message Passing Programs)
/Author (Masao Okita, Fumihiko Ino, and Kenichi Hagihara)
}

\begin{abstract}
In message passing programs, once a process terminates with an
unexpected error, the terminated process can propagate the error to the
rest of processes through communication dependencies, resulting in a program
failure. Therefore, to locate faults, developers must identify the
group of processes involved in the original error and {\em
faulty processes} that activate faults. This paper presents a
novel debugging tool, named {\em MPI-PreDebugger} (MPI-PD), for
localizing faulty processes in message passing programs. MPI-PD
automatically distinguishes the original and the propagated errors by
checking communication errors during program execution. If MPI-PD
observes any communication errors, it backtraces communication dependencies and
points out potential faulty processes in a timeline view. We also
introduce three case studies, in which MPI-PD has been shown to play
the key role in their debugging. From these studies, we believe that
MPI-PD helps developers to locate faults and allows them to concentrate in
correcting their programs.
\end{abstract}


\section{Introduction}
\label{sec:introduction}
In recent years, cluster/grid computing
\cite{buyya99cluster,ian98grid} is emerging as a cost-effective
methodology for high performance computing. The message passing
paradigm \cite{mpif94mpi} is a widely employed programming paradigm
that gives us efficient parallel programs on these computing
environments.

However, debugging message passing programs is usually time-consuming,
since we have to investigate a large amount of debugging information
compared to sequential programs. Furthermore, once a process
terminates with an unexpected error \cite{ melliar-smith77ldrs}, the
terminated process can propagate the error to the rest of processes
through communication dependencies. For example, if a process terminates
before sending an intended message, the receiver process that has no
original fault also terminates, since it fails to receive the expected
message. This {\em error propagation} makes it complicated to locate
the hidden faults from a number of observed errors.

To give developers valuable insights for debugging, a number of
debugging tools have been developed for message passing
programs. Post-mortem performance debuggers such as ParaGraph
\cite{heath91paragraph}, ATEMPT \cite{kran96graph}, XMPI
\cite{www02xmpi}, and Vampir \cite{pallas99vampir} visualize detailed
timeline view of communications, so that developers can
intuitively understand program behaviors.

Source-level debuggers such as TotalView \cite{etnus01tv}, MPIGDB
\cite{ralph00mpd}, and CDB \cite{wu02cdb} allow stepwise
execution of programs. TotalView also has a facility for visualizing,
named Message Queue Graph (MQG), which shows the states of the
pending send and receive operations. MPIGDB is based on a
sequential debugger, GDB \cite{stallman02gdb}, and allows developers
to broadcast terminal input to all GDB processes attached to
computing processes. CDB also provides a similar debugging
environment by employing GDB at its lower layer.

{\em Fault localization} \cite{jones02faultlocalzation} is another
approach for debugging programs.
{\em Relative debugging} \cite{hood00relative,gregory01relative} is
a kind of fault localization
for programs that have been
ported from sequential to parallel architectures or between different
parallel architectures.
It dynamically compares data between two
executing programs, so that can locate errors in the compared
programs. In \cite{robert96racecondition}, Netzer et al.~have pointed
out that unforeseen consequences of bugs can cause messages to arrive
in unexpected orders. Their algorithm dynamically locates errors by
detecting unintended nondeterminism, or race conditions.

Process grouping
\cite{kran02ipdps,kunz93wpdd,stringhini00pdpta} is a
fundamental technique for scalable visualizing and debugging. DeWiz
\cite{kran02pdp,kran02ipdps} aims at identifying closely related
processes and reducing the amount of trace data. Given a specific
process, DeWiz isolates the related processes according to the
accumulated length of transmitted messages.

Thus, a number of tools provide useful debugging functions. However,
developers still suffer for selecting the original error from a number
of observed errors, including original and propagated errors.
Once the original error is given to developers,
they can immediately investigate faults by using existing debuggers and
concentrate in correcting them.

In this paper, we propose a novel debugging tool, named {\em
MPI-PreDebugger} (MPI-PD), for localizing faulty processes in message
passing programs. Current MPI-PD supports programs written using the
Message Passing Interface (MPI) standard \cite{mpif94mpi} and focuses
on faults that terminate program execution. MPI-PD aims at reducing
developers' workloads required for localizing faulty processes in
timeline visualization.

To achieve this, MPI-PD dynamically checks communication errors in
accordance with the error definition in a program execution model. If
MPI-PD observes any communication errors, it then generates a trace
file, backtraces communication dependencies and points out
potentially faulty processes in a timeline view. Thus, MPI-PD reduces
the amount of debugging information before developers visualize and
investigate it by using performance debuggers and source-level
debuggers.

The rest of this paper is organized as follows. Section
\ref{sec:definition} formally characterizes communication errors in
MPI programs and makes clear the differences among faults, errors, and
failures. Section \ref{sec:algorithm} gives an algorithm for
localizing faulty processes in a given trace file while Section
\ref{sec:mpipd} presents MPI-PD, which implements the proposed
algorithm. Section \ref{sec:studies} introduces three case studies
assisted by MPI-PD. At last, Section \ref{sec:conclusions} concludes
this paper.

\section{Modeling Behavior of Message Passing Programs}
\label{sec:definition}
This section shows a definition of communication errors in MPI
programs. We define it by extending the program execution model
described in \cite{robert92model}.

\subsection{Event graph: program execution model}
\label{subsec:event_graph}
An execution of a message passing program is defined as a directed
graph, $G=(E,\rightarrow)$, where $E$ represents a finite set of
{\em events} while $\rightarrow$ represents the
{\em happened-before relation} \cite{lamport78cacm} defined over $E$
\cite{robert92model}. In the following, we call this directed graph
the {\em event graph} \cite{kran02pdp}.

An event in this context represents the execution instance of a set
of consecutively executed statements in some process
\cite{robert92model}. Any event $e \in E$ is observed during a
program execution. In the following, let $e_{p,i}$ be the
$i^{\rm th}$ event on process $p$.

The happened-before relation $\hbr$ shows how events potentially
affect one another \cite{lamport78cacm}. This relation is defined as the
irreflexive transitive closure of the union of two other relations:
$\hbr = (\sor \cup \cor)^+$. Here, $\sor$ and $\cor$ respectively
represent the sequential order relation and the concurrent order
relation as follows \cite{kran02pdp}:

\begin{figure}[ht]
\centering
\includegraphics[scale=.40]{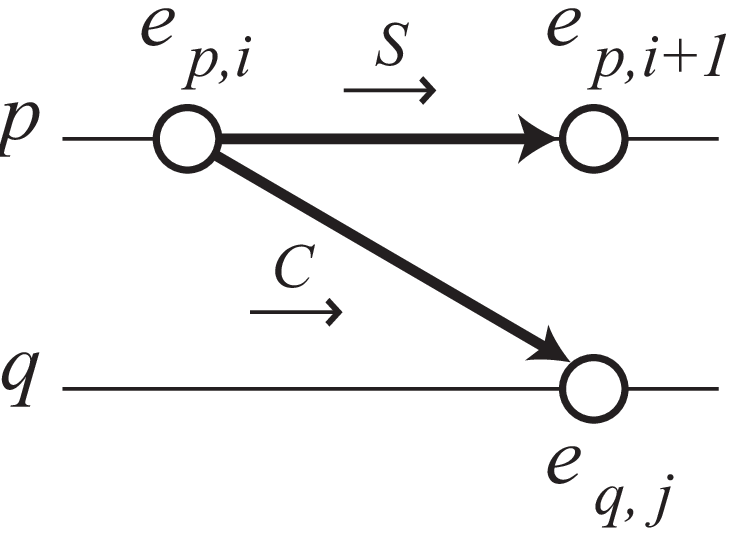}
\qquad
\includegraphics[scale=.40]{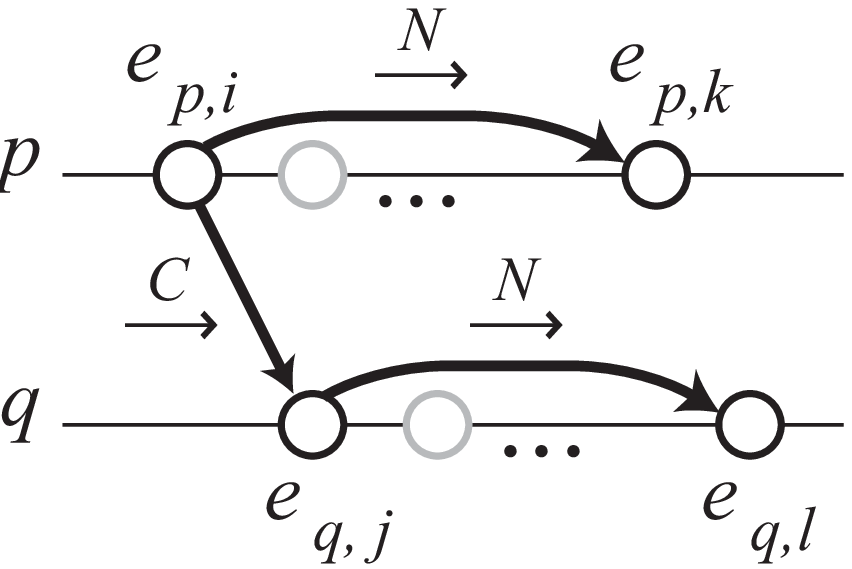}

(a) Blocking communication\hspace{1em}
(b) Nonblocking communication
\caption{Order relations between events. A node represents an event
and an arrow represents a relation.}
\label{fig:relation}
\end{figure}

\begin{description}
\item[Sequential order relation, $\sor$:]

As illustrated in Figure \ref{fig:relation}(a), the
sequential order of events, $e_{p,i} \sor e_{p,i+1}$,
defines that the $i^{\rm th}$ event $e_{p,i}$ on any
sequential process $p$ occurred before the
$i+1^{\rm st}$ event $e_{p,i+1}$.

\item[Concurrent order relation, $\cor$:]

As illustrated in Figure \ref{fig:relation}(a), the
concurrent order of events, $e_{p,i} \cor e_{q,j}$,
defines that the $i^{\rm th}$ event $e_{p,i}$ on any
process $p$ occurred directly before the $j^{\rm th}$
event $e_{q,j}$ on another process $q$, if $e_{p,i}$ is
the sending of a message by process $p$ and $e_{q,j}$ is
the receipt of the same message by another process $q$.

\end{description}

Although the event graph is a sufficient model for visualizing the
behavior of message passing programs, we have to add one relation to
this graph to characterize the errors relevant to
nonblocking communications \cite{mpif94mpi}. This additional relation
exists between a pair of events caused by the initiation and the
completion of a nonblocking send/receive operation:

\begin{description}
\item[Nonblocking order relation, $\nor$:]

As illustrated in Figure \ref{fig:relation}(b), the
nonblocking order relation, $\nor$, shows the order in
which nonblocking messages are initialized and then
completed: $e_{p,i} \nor e_{p,k}$ defines that
$e_{p,i} \sor e_{p,k}$, if $e_{p,i}$ is the send/receipt
initiation of a message by process $p$ and $e_{p,k}$ is
the completion of the same message by the same process $p$.
\end{description}

In our extended event graph, the happened-before relation is
redefined as $\hbr = (\sor \cup \cor \cup \nor)^+$.

\subsection{Fault, error, and failure}
\label{subsec:fault-error-failure}
The concepts of faults, errors, and failures
\cite{melliar-smith77ldrs} used in our discussion are briefly
explained as follows: a program with a bug has a fault in itself and
an active fault causes an error. If the error fails to be corrected,
it causes a failure.

\begin{figure}[ht]
\centering
\includegraphics[scale=.5]{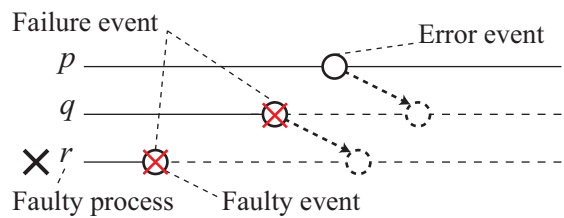}

\caption{Fault, error, and failure events. While a crossed node
represents an unexpectedly terminated event, a dotted node
represents expected but non-occurred event.}
\label{fig:fault-error-failure}
\end{figure}

Figure \ref{fig:fault-error-failure} shows an example that interprets
these three concepts on events. In this example, process $r$ is the
faulty process, since it executes a faulty statement and causes a
faulty event. It also terminates against developer's
intension, so that causes a failure event. After this, process $q$
fails to pass a message to process $r$, so that causes an error
event, resulting in a failure event (since it terminates). Process
$p$ also faces with a communication error, however, its error handler
avoids its failure.

Let $is\_failed(e)$ denote whether event $e$ causes a failure or
not. Since failure events have no successor and occur when programs
unexpectedly terminate, $is\_failed(e)$ is defined as follows:
\begin{eqnarray}
is\_failed(e) \hspace{-.5em} &=& \hspace{-.5em} {\rm~the~program~terminated~unexpectedly.} \nonumber
\label{eqn:isfailed}
\end{eqnarray}

\subsection{Communication errors in MPI programs}
\label{subsec:comm_faults}
In MPI programs, an event causes a communication error, if it
satisfies one of the following two conditions: isolated or
truncated, defined as follows;

\begin{itemize}
\item {\em Isolated events}.

\begin{itemize}
\item An event $e_{p,i}$ ($e_{q,j}$) is called an
    isolated send (receive) event,
    if $\neg \exists~e_{q,j} \in E~(e_{p,i} \in E)$
    such that $e_{p,i} \cor e_{q,j}$, respectively
    \cite{kran02pdp}.
\item An event $e_{p,i}$ ($e_{p,k}$) is called an
    isolated send/receive initiation (completion)
    event, if
    $\neg \exists~e_{p,k} \in E~(e_{p,i} \in E)$ such
    that $e_{p,i} \nor e_{p,k}$, respectively.
\end{itemize}

\item {\em Truncated events}.

\begin{itemize}
\item Two events $e_{p,i}$ and $e_{q,j}$ are called
    truncated events, if $e_{p,i} \cor e_{q,j}$ and
    $len(e_{p,i}) > len(e_{q,j})$, where $len(e_{p,i})$
    and $len(e_{q,j})$ represent the length of the send
    buffer specified in event $e_{p,i}$ and the receive
    buffer specified in event $e_{q,j}$, respectively.
\end{itemize}
\end{itemize}

Isolated events are caused under the following two situations. One is
the mismatch of occurred events and the other is the
non-occurrence of expected events. First, occurred but mismatched
events can trigger off an error propagation. For example, an MPI
routine call with an invalid tag/communicator \cite{mpif94mpi} or an
invalid source/destination rank fails to pass the
intended message. Similar mismatch can occur between the initiation
and the completion of a nonblocking send/receive operation. Next,
expected but non-occurred events cause serious problems, since they
can propagate errors through all processes. For example, if a
process terminates before sending an intended message,
the receiver process that has no original fault also terminates,
since it fails to receive the expected message. Thus, isolated events
propagate errors similarly to the domino effect,
leading to a program failure.

A pair of truncated events indicates an occurrence of an overflow at the
receive buffer. 
In a strict sense, a message should be passed between the send and the
receive operations with the same buffer length
\cite{kran02pdp}. However, as MPI does, we also permit
passing a message between events $e_{p,i}$ and $e_{q,j}$ such that
$e_{p,i} \cor e_{q,j}$ and $len(e_{p,i}) < len(e_{q,j})$. In practice,
some nondeterministic applications require this flexibility, because
the receiver processes in these applications want to receive a variable
length message at one receive operation.
Therefore, we permit passing a message
between events with different buffer length except for truncated
events.

Thus, the error of an event can depend on that of an event on another
process. In this paper we call that processes $p$ and $q$ have a
{\em communication dependency} if the error of event $e_{p,i}$ on process
$p$ determines that of event $e_{q,j}$ on another process $q$.

Here notice that MPI has four communication modes \cite{mpif94mpi}: the
standard, buffered, synchronous, and ready modes. These modes differ by
when they solve the matching of outgoing messages. For example, when
two processes send a message to each other, they fall into a deadlock
in the synchronous mode while they are deadlock-free in the buffered
mode. Therefore, we have to check communication errors without
destroying these communication semantics in the target programs. That
is, outgoing messages have to be checked in the same mode as their
original mode. The error detection mechanism employed in MPI-PD is
presented later in Section \ref{subsec:error-detection}.

For collective communications, since they can be implemented by using
point-to-point communications, we repeatedly apply the above
error definition to all of the point-to-point messages that compose
the collective communication.

In the following, let $is\_isolated(e_{p,i})$ denote whether event
$e_{p,i}$ is isolated event or not. Let
$is\_truncated(e_{p,i},e_{q,j})$ also denote whether events
$e_{p,i}$ and $e_{q,j}$ are truncated events or not.

\section{Algorithm for Localizing Faulty Processes}
\label{sec:algorithm}
This section presents the details of our proposed algorithm. We
describe how to localize faulty processes in a given event graph. We
assume here that the event graph is already generated by the error
detection mechanism presented later in Section
\ref{subsec:error-detection}.

\begin{figure}[htb]\footnotesize 
\setbox0\vbox{
1. {\bf Algorithm} LocalizeFaultyProcesses($P$, $G$, $P_e$, $E_e$)\\
\hspace{.5em}2. \hspace{1em} // Input: $P$, a set of process ranks.\\
\hspace{.5em}3. \hspace{1em} // \hspace{2.4em} $G$, an event graph.\\
\hspace{.5em}4. \hspace{1em} // Output: $P_e$, a set of localized faulty process ranks.\\
\hspace{.5em}5. \hspace{1em} // \hspace{3em} $E_e$, a set of failure events on each processes.\\
\hspace{.5em}6. {\bf begin}\\
\hspace{.5em}7. \hspace{2em}// (1) Identify failure events occurred on each processes.\\
\hspace{.5em}8. \hspace{2em}$E_e := \emptyset$;\\
\hspace{.5em}9. \hspace{2em}{\bf foreach} ($p \in P$) {\bf begin}\\
10. \hspace{4em}{\bf if} $e_{p,i}$ such that $is\_failed(e_{p,i})=true$ exists. \hspace{0.5em}{\bf then} \hspace{0.5em}$fe_p := e_{p,i}$\\
11. \hspace{4em}{\bf else}\hspace{0.5em}$fe_p := null$\\
12. \hspace{4em}{\bf endif}\\
13. \hspace{4em}$E_e := E_e \cup \{ fe_p \}$;\\
14. \hspace{2em}{\bf end}\\
15. \hspace{2em}// (2) Localize faulty processes by recursive analysis.\\
16. \hspace{2em}$P_e := \emptyset$;\\
17. \hspace{2em}{\bf foreach} ($p \in P$) {\bf begin}\\
18. \hspace{4em}{\bf if} (BacktraceCommDep($p$, $\emptyset$) $\ne$ 0)\hspace{0.5em}{\bf then}\hspace{0.5em}$P_e := P_e \cup \{ p \}$;\hspace{3em}// Process $p$ has faults.\\
19. \hspace*{2em}{\bf end}\\
20. {\bf end}\\
21. // A recursive function that backtraces communication dependencies from process $p$.\\
22. {\bf function} BacktraceCommDep($p$, $P_{dep}$)\\
23. {\bf begin}\\
24. \hspace*{2em}{\bf if} (($p \in P_e$) $\vert \vert$ (($fe_p = null$) \&\& ($P_{dep} = \emptyset$))) \hspace{0.5em}{\bf then} \hspace{0.5em}{\bf return} 0;\hspace{1.4em}// $p$ is already traced or valid.\\
25. \hspace*{2em}{\bf else if} ($fe_p$ is a calculation event) \hspace{0.5em}{\bf then} \hspace{0.5em}{\bf return} --1;\hspace{5em}// (a) Calculation fault.\\
26. \hspace*{2em}{\bf else if} ($fe_p = null$) \hspace{0.5em}{\bf then}\hspace{0.5em}{\bf return} --2;\hspace{11em}// (b) Non-occurred event.\\
27. \hspace*{2em}{\bf else if} ($p \in P_{dep}$) \hspace{0.5em}{\bf then}\hspace{0.5em}{\bf return} $p$;\hspace{12.5em}// (c) Deadlock or (d) Overflow.\\
28. \hspace*{2em}{\bf endif}\\
29. \hspace*{2em}$q := ptnr(fe_p)$;\hspace{3em}// Source/destination rank for $fe_p$\\
30. \hspace*{2em}$Q_{dep} := P_{dep} \cup \{ p \}$;\hspace{1.4em}// Update the call history.\\
31. \hspace*{2em}$retval :=$ BacktraceCommDep($q$, $Q_{dep}$);\\
32. \hspace*{2em}{\bf if} ($retval \ne 0$) \hspace{0.5em}{\bf then}\hspace{0.5em}$P_e := P_e \cup \{ q \}$;\hspace{1.4em}// Process $q$ has faults.\\
33. \hspace*{2em}{\bf if} ($retval = p$) \hspace{0.5em}{\bf then}\hspace{0.5em}$retval := 0$;\\
34. \hspace*{2em}{\bf else if} ($retval < 0$) \hspace{0.5em}{\bf then}\hspace{0.5em}$retval$++;\\
35. \hspace*{2em}{\bf endif}\\
36. \hspace*{2em}{\bf return} $retval$;\\
37. {\bf end}
}\centerline{\fbox{\box0}}
\caption{Algorithm for localizing faulty processes.}
\label{fig:algorithm}
\end{figure}

\begin{figure}[ht]
\centering
\hspace{5em}
\includegraphics[scale=.43]{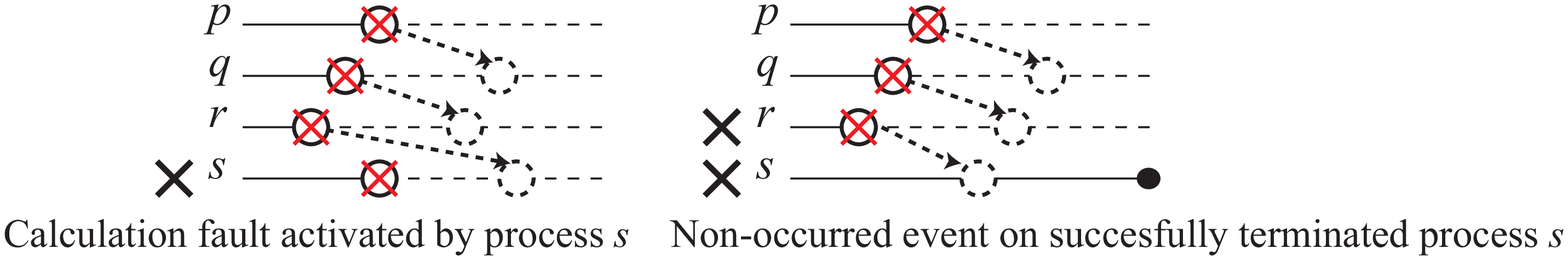}

(a) Calculation fault\hspace{4em}
(b) Non-occurred event
\includegraphics[scale=.43]{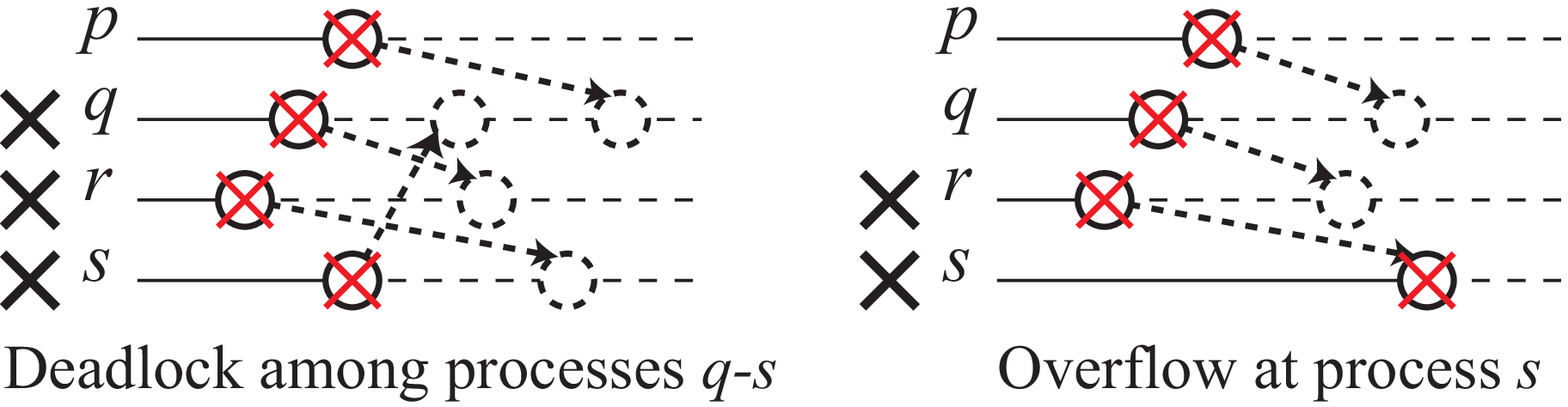}

(c) Deadlock\hspace{8em}
(d) Overflow
\caption{Four failure situations classified by proposed algorithm.}
\label{fig:failure-status}
\end{figure}

Figure \ref{fig:algorithm} shows our algorithm, which requires a set
of process ranks, $P$, and an event graph, $G$, and returns sets of
localized faulty processes and the failure events on each process,
$P_e$ and $E_e$, respectively. Our algorithm consists of two stages
as follows:

\begin{itemize}
\item Identification of failure events (see line 7--14 in Figure
\ref{fig:algorithm}).
\item Localization of faulty processes (see line 15--37 in Figure
\ref{fig:algorithm}).
\end{itemize}

At the first stage, the algorithm identifies all failure events.
After this stage, it localizes
faulty processes by backtracing communication dependencies in a recursive
manner. Our algorithm then classifies program failure into the
following four situations:

\begin{description}
\item[(a) Calculation fault:]

Figure \ref{fig:failure-status}(a) illustrates this
situation. As a result of backtracing, our algorithm finds
that process $s$ terminates unexpectedly and has no
communication dependency to any other processes. Therefore, the
algorithm determines that the faulty process is process
$s$, which causes a calculation fault.

\item[(b) Non-occurred event:]

Figure \ref{fig:failure-status}(b) illustrates this
situation, in which process $s$ has a communication
dependency from $r$ but terminates successfully.
In this situation, we think whether process $r$ could have sent a
message redundantly or process $s$ could
have missed to call a receive routine.
However, it seems to be difficult to
automatically identify the faulty process from processes
$r$ and $s$. Therefore, our
algorithm determines that the faulty processes are both of
processes $r$ and $s$, or a process left by a normally terminated
process and the terminated process.

\item[(c) Deadlock:]

A deadlock occurs if there exists a cyclic communication
dependency. In Figure \ref{fig:failure-status}(c),
processes $q$, $r$ and $s$ fall into a deadlock. Our
algorithm determines that the faulty processes are all the
processes that participate in the deadlock.

\item[(d) Buffer overflow:]

In Figure \ref{fig:failure-status}(d), process $s$ causes
a buffer overflow.
As same as situation (b), it also seems to be difficult to
identify which of processes $r$ and $s$ has called an MPI
routine with an invalid buffer length. Therefore, our
algorithm determines that the faulty processes are both of
processes $r$ and $s$, which have a pair of truncated events.
\end{description}

Notice that the algorithm described in Figure
\ref{fig:algorithm} backtraces communication dependencies by assuming that
all the source/destination ranks are valid. Therefore, if a faulty
process calls an MPI routine with an invalid source/destination,
this algorithm can omit the faulty process from the localized
processes. We discuss this problem later in Section
\ref{subsec:studies_applicability}.

\section{MPI-PreDebugger}
\label{sec:mpipd}
This section presents the details of MPI-PD, including its environment
for debugging and its mechanism for run-time error detection.

\begin{figure*}[ht]
\centering
\includegraphics[width=14.0cm]{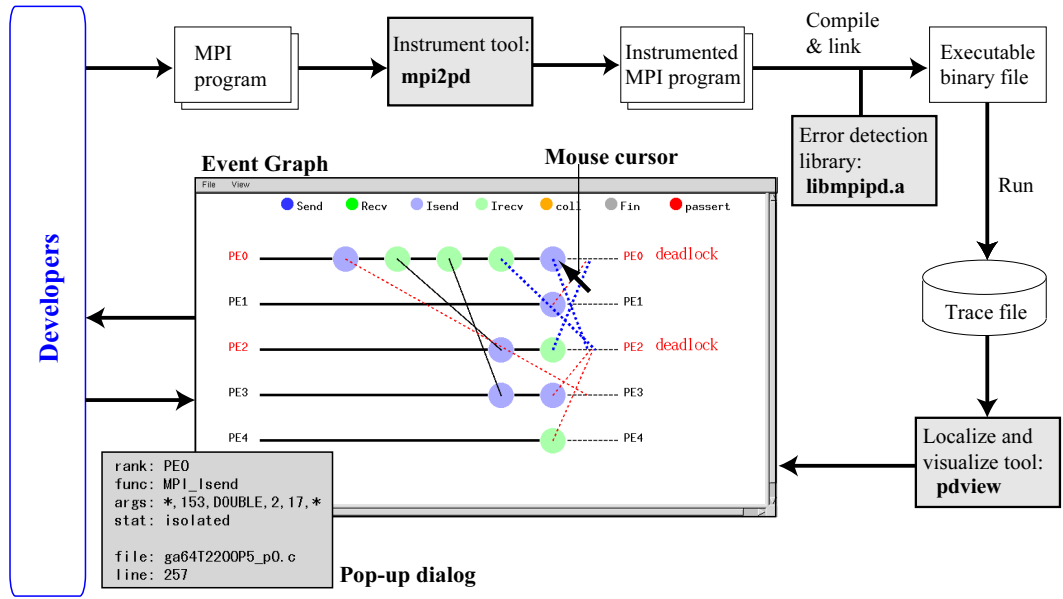}
\caption{Debugging process with MPI-PD.}
\label{fig:debugging-process}
\end{figure*}

\subsection{Overview of debugging environment}
\label{subsec:overview}

Figure \ref{fig:debugging-process} shows the debugging process with
MPI-PD. The debugging functions in MPI-PD are implemented using the
C++ language and the Ruby-GNOME toolkit \cite{www02rubygnome} and
composed of three components: the instrument tool mpi2pd, the
run-time error detection library libpdmpi.a, and the localize
and visualize tool pdview.

The instrument tool mpi2pd automatically replaces all of the
MPI routines in programs with instrumented MPI routines
based on pattern-match rules.
The instrumented routine is a combination of the original MPI routine
and the run-time error detection function.
After this replacement, developers have to generate the
object codes by compiling their programs and the executable binary
file by linking the object codes with the run-time error detection
library.

The run-time error detection library checks communication errors
whenever the processes call the instrumented MPI routines (see Section
\ref{subsec:error-detection}). If the library detects any
communication error, it terminates program execution and generates a
trace file. The trace file has the following information for every
event observed during program execution: (1) event number, (2) process
rank, (3) corresponding line in source code and its file name, and (4)
corresponding MPI routine and its arguments.

Given a trace file, the visualization tool pdview allows developers to
view the behavior of the terminated program, as shown in Figure
\ref{fig:debugging-process}. It visualizes the event graph, which
has the process axis in vertical and the time axis in horizontal,
and shows the result of the fault localization described in Section
\ref{sec:algorithm}. In the event graph, a colored node corresponds
to an event and the type of the MPI operation that caused the event
decides its color. A solid line between two nodes
corresponds to a successful communication while a dotted line
corresponds to a failure communication.

In default mode, pdview avoids visualizing the entire event graph. It
visualizes all of failure events occurred on each process and the
successful events occurred directly before the failure
events. Furthermore, pdview can isolate faulty processes from
the event graph. Developers can visualize an isolated event graph by
selecting process whichever they want. In addition to
these visualization functions, pdview also shows following information:

\begin{itemize}
\item Faulty processes localized by the proposed algorithm.
\item Failure situation selected from four situations (see
Figure \ref{fig:failure-status}).
\end{itemize}

Furthermore, developers can investigate every visualized event. If
they click the mouse on a node in the visualized event graph, then
pdview pops up a dialog, which shows information (1)--(4) about the
corresponding event and its error reason (isolated/truncated). This
information is useful for developers to locate faults in
programs. After this fault localization, source-level debuggers can
effectively assist developers to investigate the detailed behavior of
the localized part.

\subsection{Mechanism for run-time error detection}
\label{subsec:error-detection}
MPI-PD checks the occurrence of communication errors during program
execution. If it detects any errors, it generates a trace file.

To realize this, we employ three methodologies. We first discuss on
the synchronous blocking send ({\tt MPI\_Ssend}) then others. The
three methodologies are as follows:

\begin{itemize}
\item Manager process: 
To generate trace files under a deadlock situation, we
employ a manager process $M_p$ for every process $p$. 
$M_p$ checks the value
of $is\_failed(e_{p,i})$ before its responsible process $p$
executes event $e_{p,i}$.
We present later how to check $is\_failed(e_{p,i})$
at next paragraph.
If $M_p$ obtains
$is\_failed(e_{p,i})=false$, it allows $p$ to execute event
$e_{p,i}$ and pushes the information about $e_{p,i}$ into
its local Event Graph $E_p$. Otherwise, it detects a
communication error, terminates $p$ and generates a trace file
from $E_p$.

\item Message queue: To handle nonblocking communications, we employ
a message queue. For nonblocking communications, to decide the
failure of completion event $e_{p,k}$, we have to refer the
information about its corresponding initiation event $e_{p,i}$
($e_{p,i} \nor e_{p,k}$). Therefore, for all processes $p$,
manager $M_p$ has its own message queue $Q_p$ for
referring to the information about the past events.

\item Timeout mechanism: We also employ a timeout mechanism due to
the difficulty in distinguishing the valid and the invalid
computation. For example, a receive event $e_{q,j}$ that never
receive a message has to be decided as
$is\_isolated(e_{q,j})=true$. However, it is hard for $M_q$
to identify whether the sender $p$ sends the message or not.
That is, $p$ can send the message after heavy computation or can
fall into an infinite loop. Therefore,
$M_p$ holds a timeout time $t(e_{p,i})$ for every
$e_{p,i}$ and decides $is\_isolated(e_{p,i})=true$ when the
time is up.
\end{itemize}

Figure \ref{fig:error-detection} shows the process of run-time error
detection for {\tt MPI\_Ssend}. In Figure \ref{fig:error-detection},
the manager of the sender has three states (states C, S1 and S2) and
that of the receiver has four states (states C, R1, R2 and R3) as
follows:

\begin{figure*}[ht]
\centering
\includegraphics[width=6.0cm]{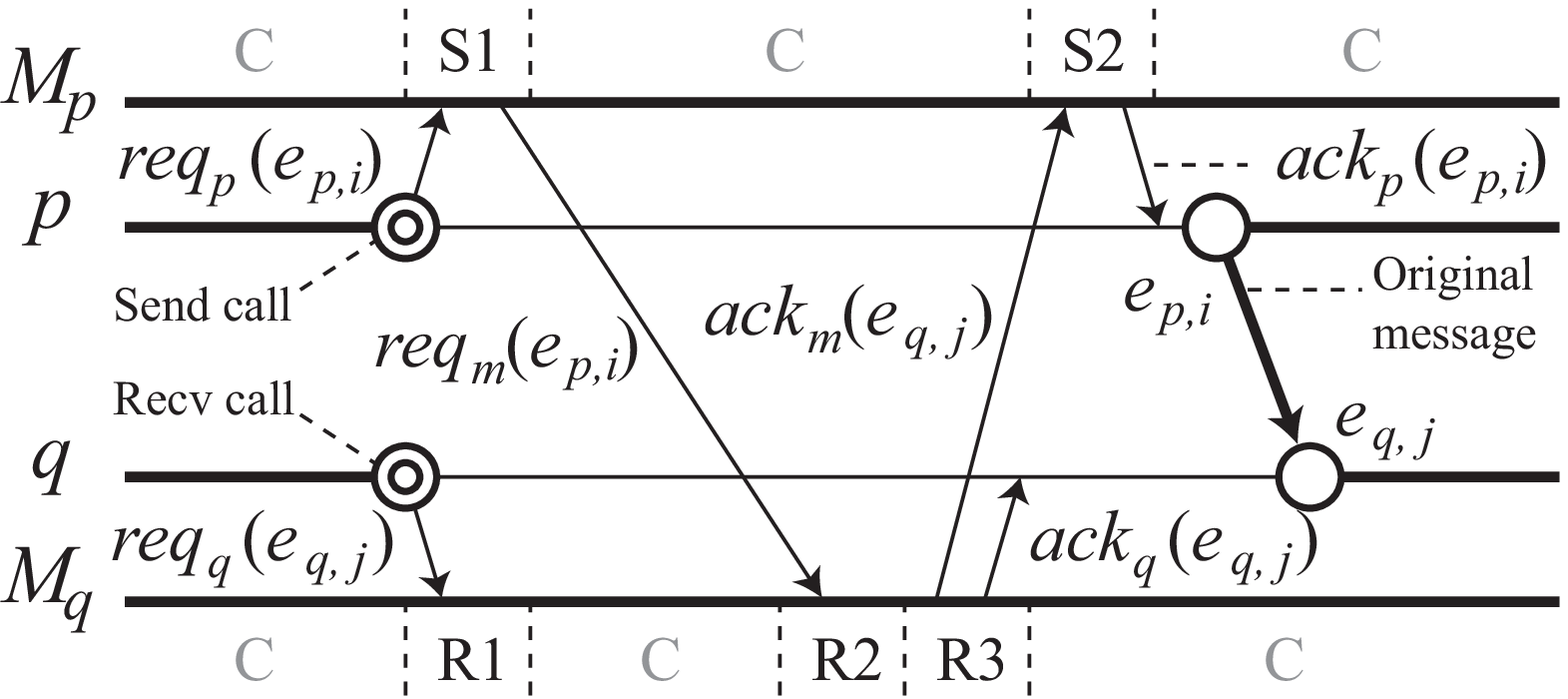}
\qquad
\includegraphics[width=6.0cm]{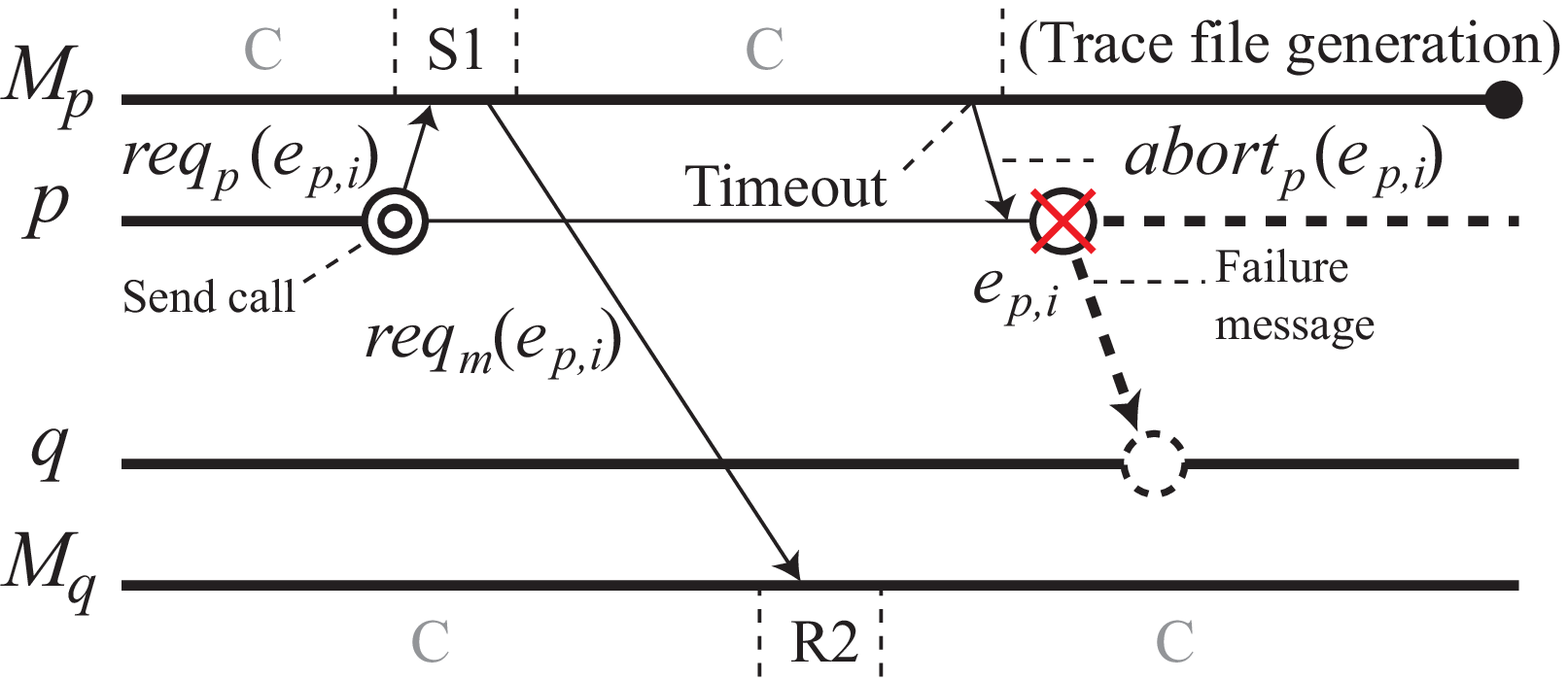}

(a) Successful case \hspace{13em}(b) Failure case
\caption{Process of run-time error detection for the synchronous
blocking send ({\tt MPI\_Ssend}). Events $e_{p,i}$ and $e_{q,j}$
correspond to {\tt MPI\_Ssend} and {\tt MPI\_Recv} calls,
respectively.}
\label{fig:error-detection}
\end{figure*}

\begin{description}
\item[Common state for the sender/receiver:]
\item State C: {\em Timeout checking and control-message waiting}. In
this state, $M_p$ continues to check $Q_p$ whether there
exist any timeout events, until it receives any control
message (ack or request messages) from $p$ or another
manager. If $M_p$ detects a timeout event $e_{p,i}$,
then it decides $is\_failed(e_{p,i})=true$ and sends an
abort request $abort_p(e_{p,i})$ to $p$. It also adds the
failure event $e_{p,i}$ to $E_p$ and terminates.
If $M_p$ receives a control message, then it changes
its state to an appropriate state.

\item[States for the sender:]
\item State S1: {\em Send initiating}. If $M_p$ receives a send
request $req_p(e_{p,i})$ from $p$, then it pushes the
information about $e_{p,i}$ into $Q_p$ with
$t(e_{p,i})$. It also checks the destination rank of
$e_{p,i}$ and transmits a send request $req_m(e_{p,i})$ to
the destination process's manager, $M_q$ (go to state C).
\item State S2: {\em Message sending}. If $M_p$ receives an
ack $ack_m(e_{q,j})$ from another manager, then it
searches $Q_p$ and selects
$e_{p,i}$ such that $is\_isolated(e_{p,i}) = false$. It
also checks whether $e_{p,i}$ and $e_{q,j}$ are
truncated events.
\begin{itemize}
\item If $is\_truncated(e_{p,i},e_{q,j})=false$, $M_p$
     decides $is\_failed(e_{p,i})=false$ and sends an
     ack $ack_p(e_{p,i})$ to
     $p$. After this acknowledgement, it deletes
     $e_{p,i}$ from $Q_p$, and adds
     both $e_{p,i}$ and $e_{q,j}$ to $E_p$ (go to
     state C).
\item Otherwise, $M_p$ decides
     $is\_failed(e_{p,i})=true$ and sends an abort
     request $abort_p(e_{p,i})$ to $p$.
     It also adds both $e_{p,i}$ and $e_{q,j}$ to $E_p$
     as failure events and terminates.
\end{itemize}

\item[States for the receiver:]
\item State R1: {\em Receive initiating}. If $M_q$ receives a receive
request $req_q(e_{q,j})$, it then searches $Q_q$ and
selects $e_{p,i}$ such that $is\_isolated(e_{p,i}) \lor
is\_isolated(e_{q,j})=false$.

\begin{itemize}
\item If such $e_{p,i}$ exists, $M_q$
     decides that $e_{p,i}$ and $e_{q,j}$ are the
     matching events (go to state R3).
\item Otherwise, it leaves the error detection on
     $e_{q,j}$ and pushes the information about
     $e_{q,j}$ into $Q_q$ with $t(e_{q,j})$ (go to
     state C).
\end{itemize}

\item State R2: {\em Send-request receiving}. If $M_q$
receives a request $req_m(e_{p,i})$ from another manager,
then it searches $Q_q$ and selects
$e_{q,j}$ such that
$is\_isolated(e_{p,i}) \lor is\_isolated(e_{q,j})=false$.

\begin{itemize}
\item If such $e_{q,j}$ exists, $M_q$
     decides that $e_{p,i}$ and $e_{q,j}$ are the
     matching events (go to state R3).
\item Otherwise, it leaves the error detection on
     $e_{p,i}$ and pushes the information about
     $e_{p,i}$ into $Q_q$ with $t(e_{p,i})$ (go to
     state C).
\end{itemize}

\item State R3: {\em Message receiving}. $M_q$ sends an ack
$ack_m(e_{q,j})$ to $M_p$.  It then checks if
$e_{p,i}$ and $e_{q,j}$ are truncated events.

\begin{itemize}
\item If $is\_truncated(e_{p,i},e_{q,j})=false$, then
     $M_q$ decides $is\_failed(e_{q,j})=false$
     and sends an ack $ack_r(e_{q,j})$ to
     $q$. After this acknowledgement, it deletes
     $e_{q,j}$ ($e_{p,i}$) from $Q_q$
     and adds both $e_{p,i}$ and $e_{q,j}$ to $E_q$
     (go to state C).
\item If $is\_truncated(e_{p,i},e_{q,j})=true$, then
     $M_q$ decides $is\_failed(e_{q,j})=true$
     and sends an abort request $abort_q(e_{q,j})$ to $q$.
     It also adds both $e_{p,i}$ and $e_{q,j}$ to $E_q$
     as failure events and terminates.
\end{itemize}
\end{description}

The manager processes buffer all events until they detect an error,
so that their local memory are possibly full.
Our algorithm described in Figure \ref{fig:algorithm} requires
failure events on each process.
Therefore, if local memory of $M_p$ is full,
we allow $M_p$ to delete information about the oldest successful event 
from $E_p$.

Here, recall that we have to keep the communication semantics, as
explained in Section \ref{subsec:comm_faults}. Therefore, for the
blocking buffered mode send ({\tt MPI\_Bsend}), we alter the sequence
of error detection. That is, to keep the buffered behavior of message
passing, process $p$ passes the original message immediately after
sending request $req_p(e_{p,i})$ to its manager $M_p$. This
alternation omits receiving an ack $ack_p(e_{p,i})$ from
$M_p$. Instead of this omission, $p$ checks an abort message
$abort_p(e_{p,i})$ from $M_p$ whenever it calls an
instrumented MPI routine. If $p$ receives the abort message
$abort_p(e_{p,i})$, it terminates its execution. Otherwise, it
continues processing the original routine.
This alteration allows $p$ to execute a few events after an original
faulty event, however there is no influence on faulty process
localization since $M_p$ identifies the faulty event correctly.

For nonblocking communications, we process states S1 and R1 at the
send initiation and the receive initiation of nonblocking operations,
respectively; and process send acks at the completion of the nonblocking
operations. For collective communications, we can apply the same
approach as for the blocking mode point-to-point routines, since the
collective communications can be implemented by using those
point-to-point routines.

Thus, exchanging information about every event among managers
enables us to detect communication errors and generate trace files
before program failure.

\section{Case Studies: Debugging Message Passing Programs with MPI-PD}
\label{sec:studies}

In this section we introduce three case studies. The aim of each
study is to investigate the effectiveness of MPI-PD from the
following point of view:

\begin{enumerate}
\item {\em Applicability}:
We investigated what kinds of faults are effective for MPI-PD.
To do this, we applied MPI-PD to a few ten of the Gaussian
programs developed by MPI beginners (see Section
\ref{subsec:studies_applicability}).

\item {\em Scalability}:
This study shows an example of scalable debugging using
MPI-PD. We applied MPI-PD to a parallel rendering program
\cite{takeuti03sac} developed by MPI experts on 64 processes
(see Section \ref{subsec:studies_scalability}).

\item {\em Usability}:
We investigated the usability of faulty process
localization. To do this, we applied MPI-PD to a complicated
program generated automatically by a parallelizing
compiler \cite{y-yamamt01ebcsh}. We also compared
visualization results between proposed MPI-PD and existing
TotalView \cite{etnus01tv} (see	Section
\ref{subsec:studies_usability}).
\end{enumerate}

\begin{table*}[tb]
\caption{Summary of case studies. $|L|$, $|P|$, and $|E|$ represent
the numbers of lines, processes, and events, respectively.}
\label{tab:summary}
\begin{center}
\begin{tabular}{|l|l|c|l|c|c|}\hline
\lw{Case study}  & \multicolumn{3}{l|}{Details of program}
      & \multicolumn{2}{l|}{Details of trace file}
      \\ \cline{2-6}
     & Developer & $|L|$ & Employed MPI routines
      & $|P|$ & $|E|$
      \\ \hline
1. Applicability & Beginner  & ~~~300 & {\tt Send}, {\tt Recv}, {\tt Isend}, {\tt Irecv}, {\tt Wait}
      & ~~~4~~ & ~~412
      \\ \hline
2. Scalability   & Expert    & 40,000 & {\tt Send}, {\tt Recv}, {\tt Sendrecv}
      & ~~64~~ & 9,774
      \\ \hline
3. Usability     & Compiler   & 20,000 &{\tt Isend}, {\tt Irecv}, {\tt Waitall}
      & ~~15~~ & ~~253
      \\ \hline
\end{tabular}
\end{center}
\end{table*}

Table \ref{tab:summary} shows a summary of the above studies. In the
following, we omit ``{\tt MPI\_}'', the prefix of MPI routines, as shown
in Table \ref{tab:summary}.

In these studies we used a PC cluster with 64 symmetric
multiprocessor (SMP) nodes. Each node in the cluster has two Pentium
III 1GHz processors and connects to a Myrinet-2000
switch \cite{nanette95myrinet}. We also employed an MPI
implementation, MPICH-GM \cite{www02mpichgm}.

\subsection{Study 1: Applicability of MPI-PD}
\label{subsec:studies_applicability}

In this study, we applied MPI-PD to 28 faulty programs developed by
six graduate students through a practice in MPI programming. These
programs solve simultaneous equations using Gaussian elimination.

\begin{table}[tb]
\caption{Application results of MPI-PD.}
\label{tab:application}
\begin{center}
\begin{tabular}{|l|c|c|}\hline
\lw{Debugging phase}       & \multicolumn{2}{c|}{Number of programs}\\ \cline{2-3}
		& Success & Failure\\ \hline
MPI Program execution      & 13 of 28   & 15 of 28\\ \hline
Event graph visualization  & 15 of 15   & \hspace{.5em}0 of 15 \\ \hline
Faulty process localization & 12 of 15   & \hspace{.5em}3 of 15\\ \hline
\end{tabular}
\end{center}
\end{table}

We first executed the programs on our PC cluster and then
visualized localization results by using MPI-PD. Table
\ref{tab:application} shows the application results at each
debugging phase.

At the execution phase, 15 of 28 programs unexpectedly terminated. As
we mentioned in Section \ref{sec:introduction}, since current MPI-PD
focuses on faults with program failures, it failed to visualize the
event graph for the remaining 13 programs that never terminated but
returned incorrect results. These programs contain semantic faults
such as invalid specifications of operators/variables and invalid
writing to message buffers before the completion of nonblocking
communications.

At the localization phase, MPI-PD successfully localized faulty
processes for 12 of 15 programs while it failed to localize them for
the remaining three programs. These three programs have calculation
faults activated by all processes at the same statement. Therefore,
every process terminated outside the instrumented MPI routines, so
that their trace files contained no information about failure
events. Thus, MPI-PD failed to localize their faulty
processes. However, in these cases, since every process terminates
without any communication dependency, error propagation is unable to
occur. Therefore, developers have to investigate every process. That
is, they have to investigate their programs between the last MPI
routine executed in a success and the next MPI routine expected to be
executed, especially where the common statements that every process
executes.

The 12 programs which MPI-PD successfully localized had a variety
of faults classified into following four types.
Notice that MPI-PD localized not the faults but the faulty processes
which activate them.

\begin{itemize}
\item Invalid source/destination rank (six programs).
\item Invalid length of message buffer (three programs).
\item Calculation fault (two programs).
\item Deadlock occurred when passing long messages (one program).
\end{itemize} 

We next confirmed that there was no faulty process omitted from the
localized results. For all cases where invalid source/destination
ranks were specified, MPI-PD pointed out deadlock processes,
including the faulty process. Therefore, the deadlock processes
pointed out by MPI-PD can include valid processes, so that there
exists a room for improving the accuracy of localization. However,
this redundancy was a little problem for the programs applied in this
study. Since their faults appear on any number of processes,
developers are allowed to scale down the number of processes without
missing the activated faults.

\subsection{Study 2: Scalable debugging with MPI-PD}
\label{subsec:studies_scalability}

\begin{figure}[ht]
\centering
\includegraphics[width=8.3cm]{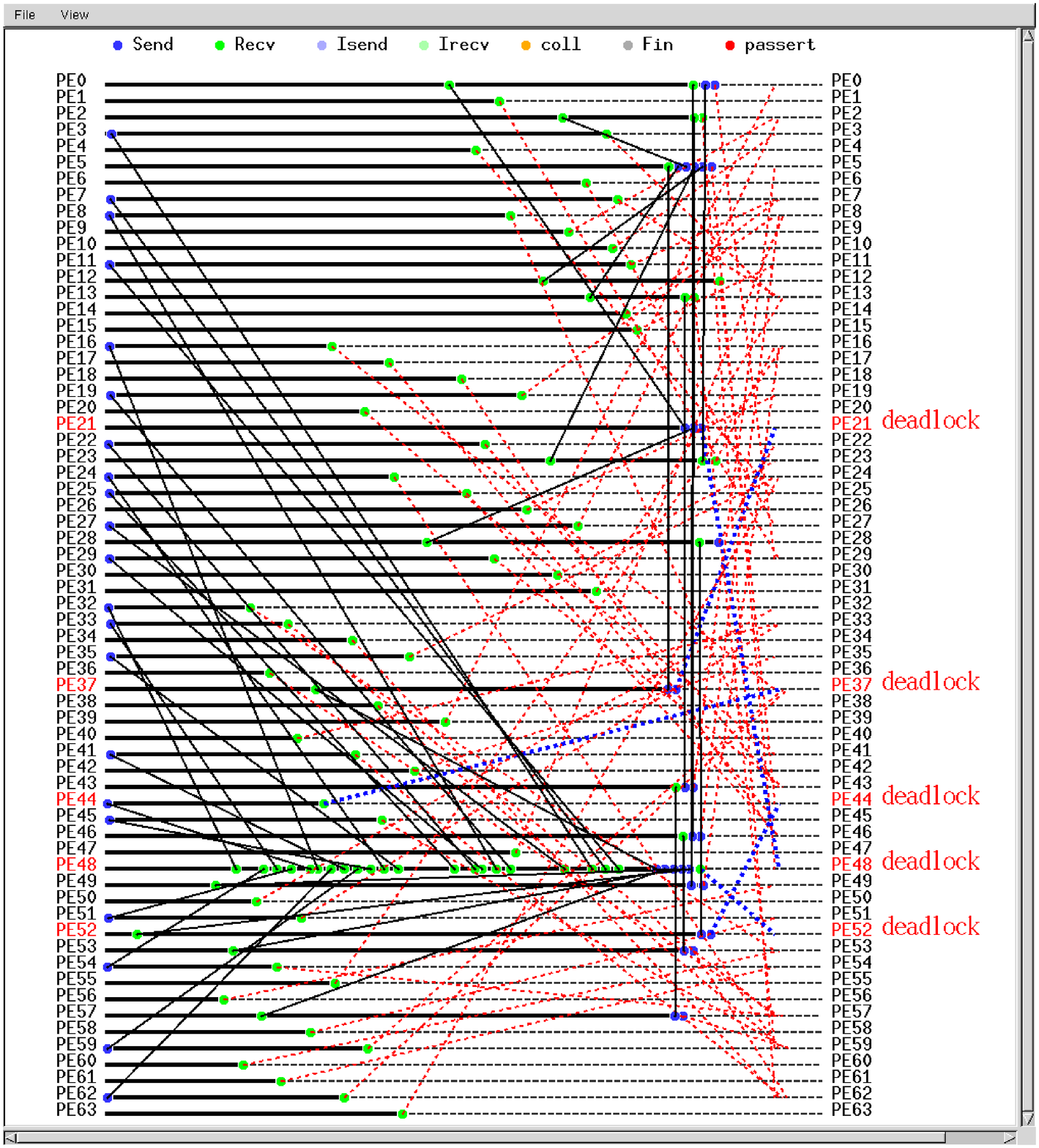}
\caption{Localized faulty processes in event graph visualized by MPI-PD.}
\label{fig:vr-all}
\end{figure}

We applied MPI-PD to a parallel rendering program
\cite{takeuti03sac} implemented on 64 processes. This program has a
fault in gathering and compositing rendered images generated by distributed
processors. For the purpose of high-speed compositing, the
developers have implemented own collective communication routines
for the gather and the broadcast operations by using point-to-point
routines, {\tt Send} and {\tt Recv}. Their collective routines are
called at every compositing stage with splitting the processes into
two groups. That is, given $n$ processes, each of $2^{i-1}$ groups
performs collective communications at the $i^{\rm th}$ stage, where
$1 \leq i \leq \log n$.

Figure \ref{fig:vr-all} shows the event graph for all processes
visualized by MPI-PD. While the program generates the total of 9,774
events, the visualized event graph is composed of 164 events classified into 64
failure events and 100 successful events occurred directly before the
failure events. In Figure \ref{fig:vr-all}, MPI-PD points out five
faulty processes from 64 processes: processes PE21, PE37, PE44, PE48,
and PE52. It also points out that these five processes fall into a
deadlock and that each of them has one failure event.

\begin{figure}[ht]
\centering \includegraphics[width=8.3cm]{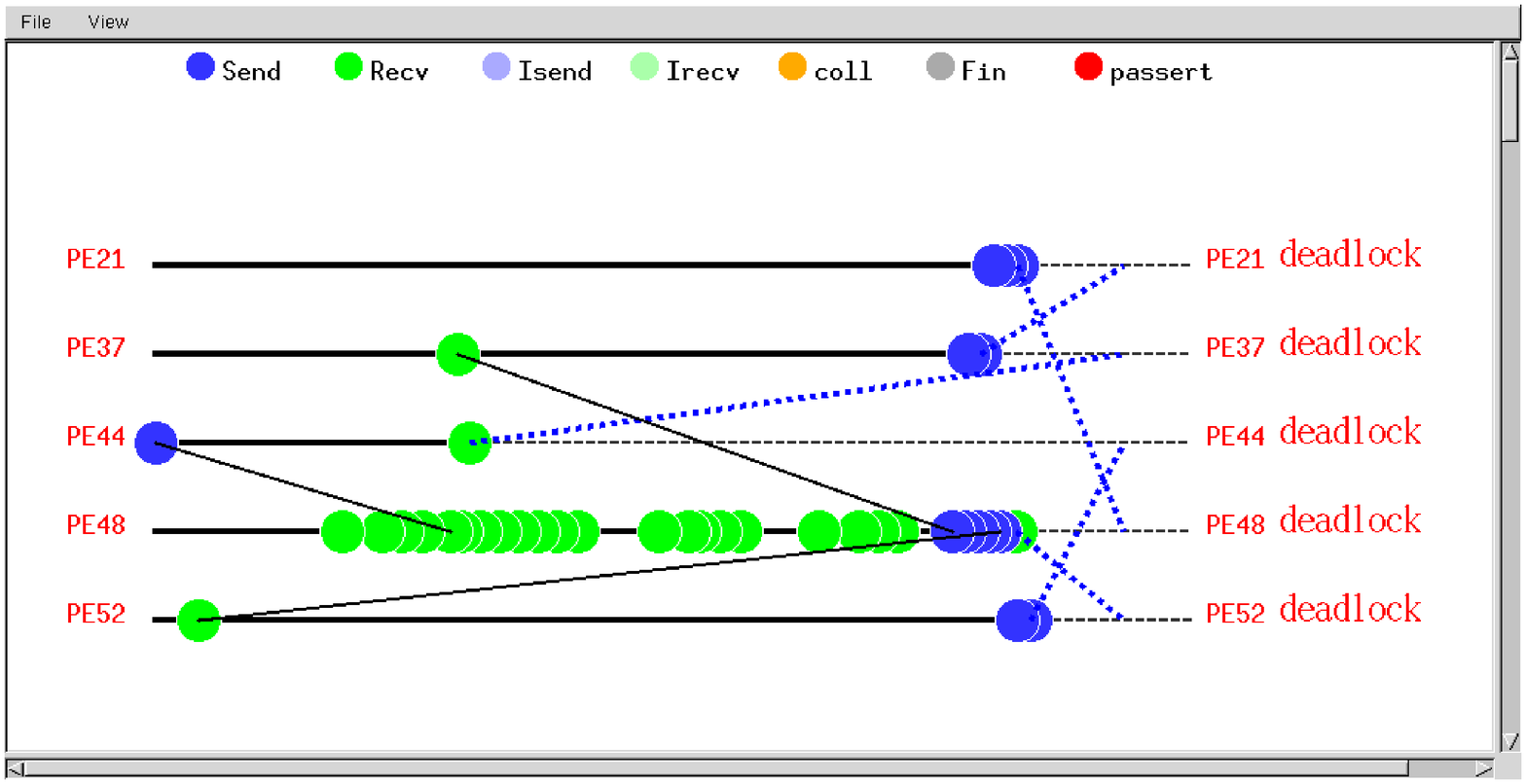}
\caption{Faulty processes isolated by MPI-PD. This graph shows only
faulty processes and communications among them.}
\label{fig:vr-fp}
\end{figure}

\begin{figure}[ht]
\centering \includegraphics[width=8.3cm]{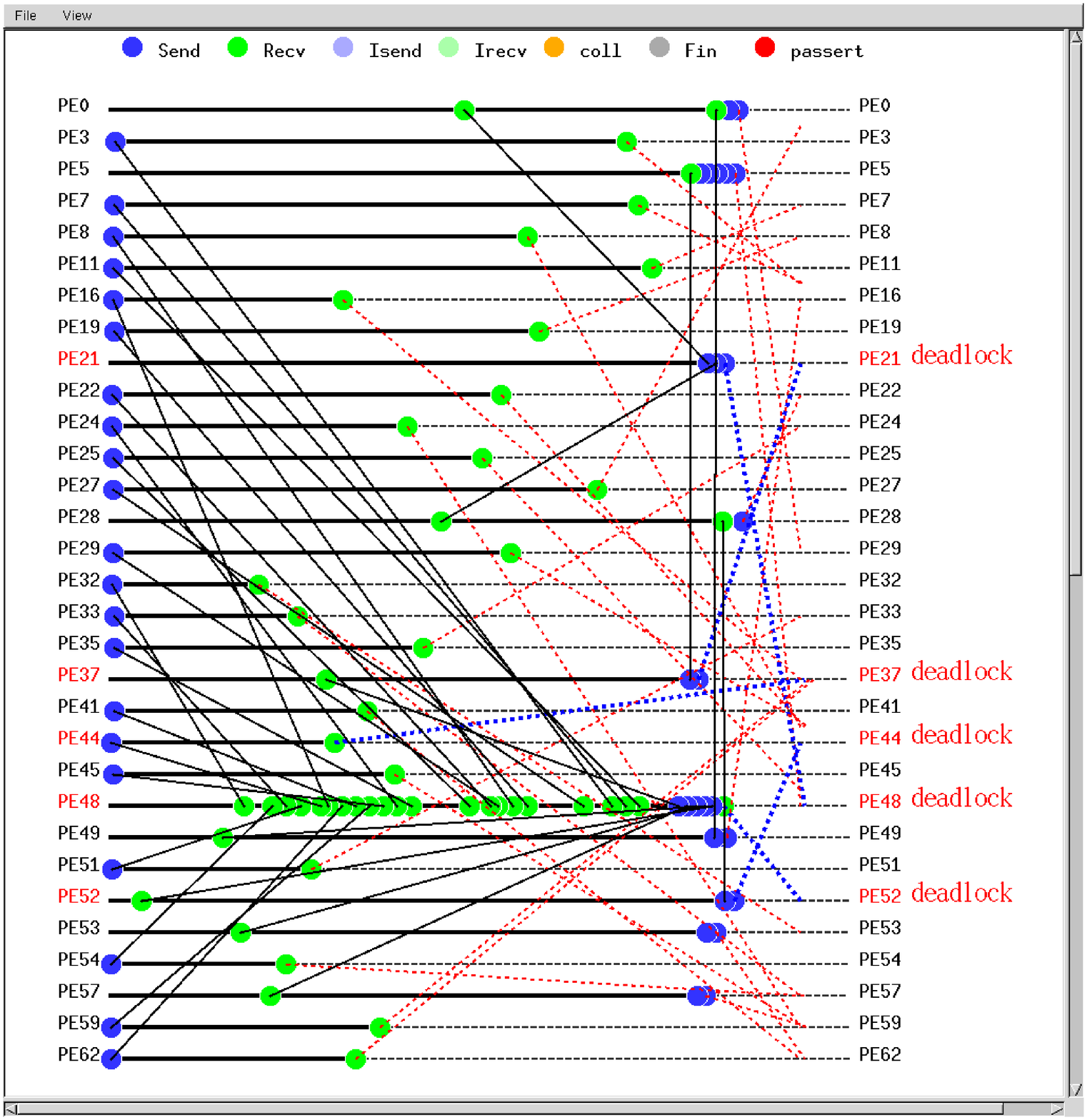}
\caption{Faulty processes and their related processes isolated by MPI-PD.
Related processes are such that faulty processes communicate with them.}
\label{fig:vr-fpplus}
\end{figure}

As we mentioned in Section \ref{subsec:overview}, MPI-PD allows
developers to visualize specific processes whichever they want. For
example, developers can view only the deadlock processes as shown in
Figure \ref{fig:vr-fp}, so that easily know how the processes fell
into the deadlock. They can also add related processes that
communicated to the deadlock processes (see Figure
\ref{fig:vr-fpplus}), so that intuitively know process PE48
received many messages compared to the other four faulty processes:
processes PE21, PE37, PE44, and PE52.

Thus, MPI-PD guided the developers to the five faulty events, so that
they easily found that process PE48, the root process of a
broadcast operation, called an excessive {\tt Send}
routine due to the lack of a {\tt break} statement. Therefore, MPI-PD assists
developers in scalable debugging, where the numbers of processes and
events are too large for them to understand the behavior of programs.

We also indicate that the buffered send operation makes it complicated
to locate faults, since this operation causes a gap between the faulty
send event and the failure event. For example, when we executed
the rendering program without error detection, since process PE48
pushed out messages in the buffered mode, it successfully returned
from the faulty {\tt Send} routine and terminated at a succeeding {\tt
Recv} routine. Therefore, without MPI-PD, the developers can
investigate the {\tt Recv} routine, which causes a non-original fault,
or a fault due to error propagation. Thus, MPI-PD's run-time error
detection is necessary for handling the buffered send operation.

\subsection{Study 3: Comparison with existing debuggers}
\label{subsec:studies_usability}

To make clear the usability of fault localization, we compared MPI-PD
with TotalView \cite{etnus01tv} by applying them to a complicated program. This
program is automatically generated by a parallelizing compiler based
on a task scheduling algorithm, Scheduling with Packaged
Point-to-point Communications (SPPC) \cite{y-yamamt01ebcsh}.

The MPI program generated by SPPC consists of two layers, the calculation
and the communication layers, which repeatedly appear during program
execution. In the calculation layer, each process independently
performs calculation without any communication. In the communication
layer, it exchanges messages by calling nonblocking communication
routines. Each process first calls many initiation routines,
{\tt Isend} and {\tt Irecv}, then a completion routine, {\tt
Waitall}. Since the parallelizing compiler mechanically generates
large-scale MPI programs, it requires a complicated work to debug
them. Furthermore, since the {\tt Waitall} routine completes all of
initiated communications at a time, it is time-consuming to
distinguish failure communications from a number of communications
completed by the {\tt Waitall} routine.

\begin{figure}[ht]
\centering
\includegraphics[width=6.0cm]{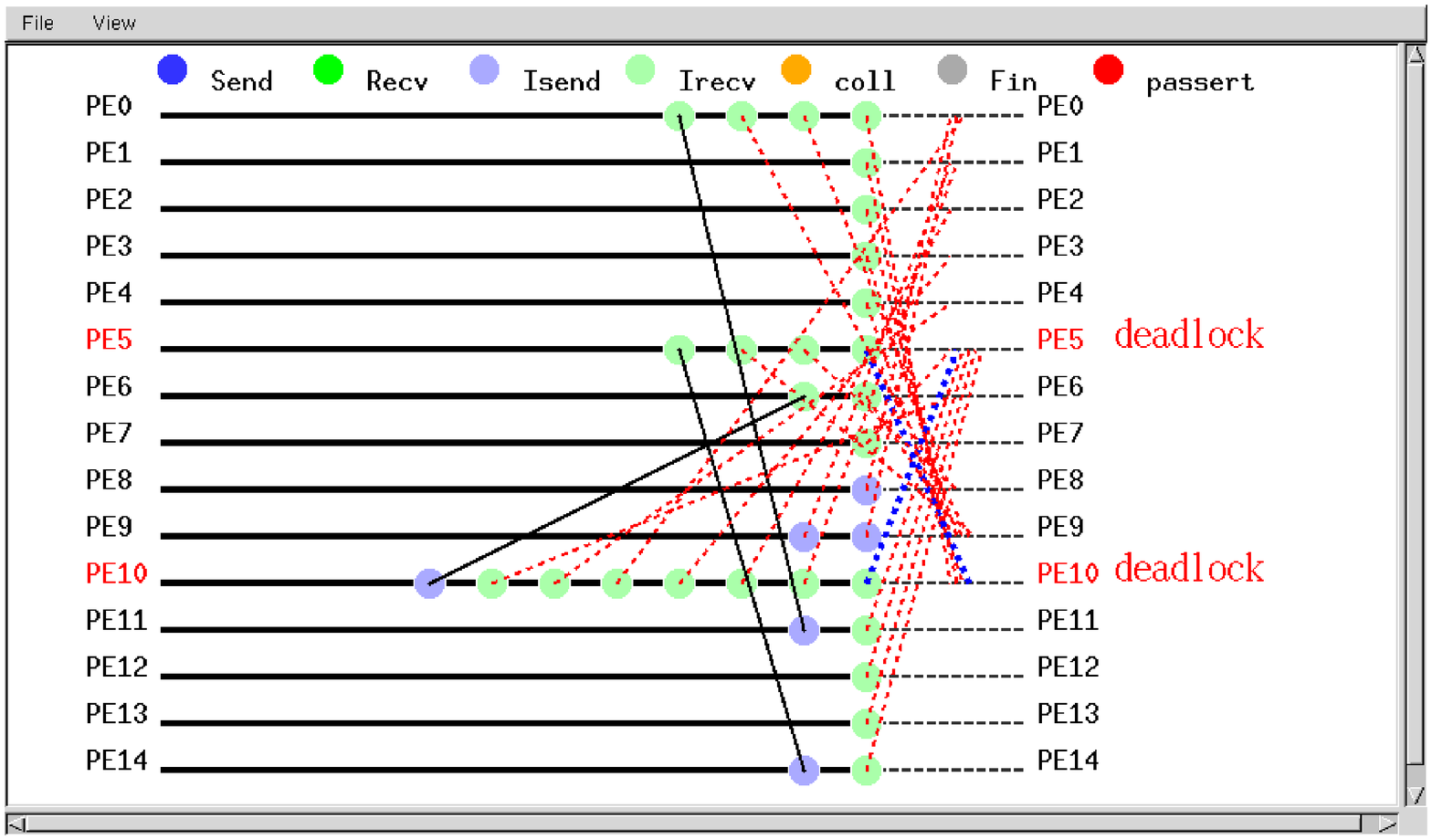}
\qquad
\includegraphics[width=6.0cm]{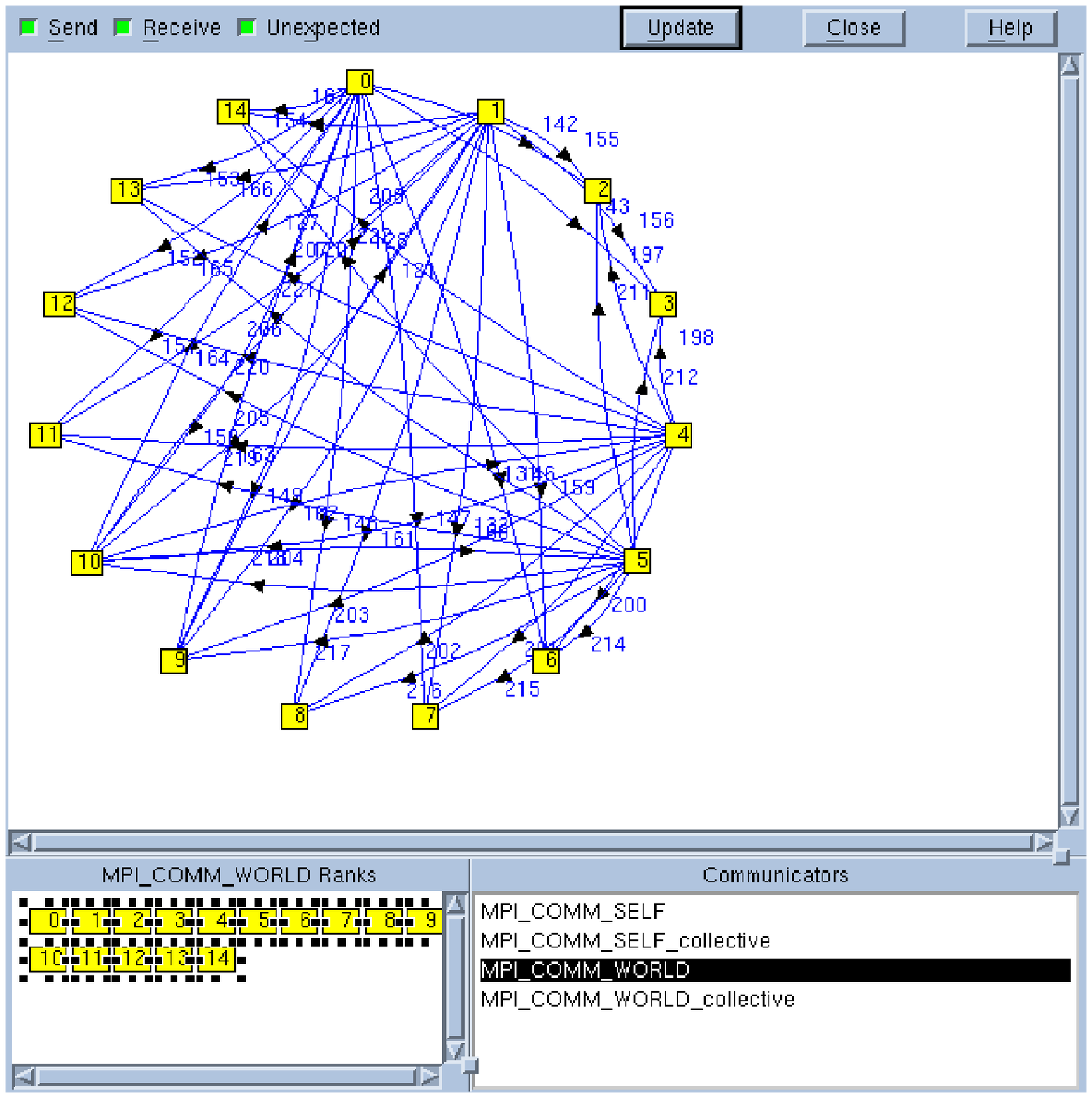}

\hspace{1em}(a) Event Graph by MPI-PD \hspace{6em}(b) Message Queue Graph by TotalView
\caption{Visualizations obtained by MPI-PD and TotalView.}
\label{fig:sppc}
\end{figure}

Figure \ref{fig:sppc} shows the visualizations
obtained by MPI-PD and TotalView. While MPI-PD
visualizes all of failure events occurred on each process and the
successful events occurred directly before the failure events,
TotalView shows {\em pending sends/receives} and {\em
unexpected messages} \cite{james99debugger,etnus01tv} at an arbitrary execution step.
Pending sends/receives represent the sends/receives that have been
initiated but have not yet been matched. Unexpected messages
represent messages that have been sent to a process but have not
yet been received.

In this program, every process terminated at a call of {\tt Waitall}
routine. At the termination, the processes tried to complete the total
of 171 nonblocking operations. For this faulty program, TotalView
visualizes 50 pending receives, represented as arrows in Figure
\ref{fig:sppc}(b). However, it is time-consuming for the developers to
investigate each of the 50 pending receives. On the other hand, MPI-PD
checks the error of every communication and localizes faulty
processes, so that it visualizes 34 of 171 events as shown in Figure
\ref{fig:sppc}(a). Since eight of 34 events are successfully communicated
events, MPI-PD reduces the number of events that have to be
investigated from 171 to 26 events. Furthermore, it points out that
processes PE5 and PE10 fall into a deadlock. Here, processes PE5 and
PE10 have three and seven error events, respectively, so that the
number of events that have to be investigated is reduced further from
171 to 10 events.

With the assistance of MPI-PD, the developer has successfully debugged
this program less than five minutes. He first investigated process PE5
and confirmed that it had no fault, and then process
PE10. At last, he reached at the fault where an invalid source was
specified at an {\tt Irecv} routine.

\begin{table*}[tb]
\caption{Difference among MPI-PD, TotalView, and DeWiz.}
\label{tab:difference}
\begin{center}
\begin{tabular}{|l|c|c|c|} \hline
Function                        & MPI-PD & DeWiz \cite{kran02pdp,kran02ipdps} & TotalView \cite{etnus01tv}\\ \hline
1. Faulty process localization & by dependency analysis &   --- & ---\\ \hline
2. Run-time error detection     & every message & every message & every message\\ \hline
3. Process grouping             & by dependency analysis & by message length & ---\\ \hline
4. Timeline visualization       &    yes &   yes & ---\\ \hline
5. Trace file reduction         &    --- &   yes & ---\\ \hline
6. Stepwise execution           &    --- &   --- & yes\\ \hline
\end{tabular}
\end{center}
\end{table*}

Table \ref{tab:difference} summarizes the difference among MPI-PD,
TotalView, and DeWiz \cite{kran02pdp,kran02ipdps}. While MPI-PD is
useful to reduce events that have to be investigated, TotalView allows
us to execute the target program in stepwise. DeWiz also provides an
analysis using the event graph. However, DeWiz aims at
identifying closely related processes and reducing the total amount of
trace data. In DeWiz, by giving a specific process, then its process
grouping function accumulates the length of transmitted messages for
every pair of processes and isolates related processes by using a
certain threshold. Therefore, developers have to decide which
processes have to be specified, and this is a similar problem addressed in
this paper. Furthermore, since error propagation has no relevance to message
length, their message length based approach is inappropriate for the
purpose of faulty process localization.

Summarizing the above discussions, DeWiz is useful to reduce the
total amount of trace files and TotalView is useful to investigate
the detailed behavior of programs. MPI-PD is useful to reduce the
number of events that have to be investigated for
debugging. Therefore, we think that appropriate combined use of these
tools is a good choice for debugging message passing programs.
For example, we first localized faulty processes by using MPI-PD 
and next investigate them in detail by using TotalView.

\section{Conclusions}
\label{sec:conclusions}
 We have presented a novel debugging tool, named MPI-PD, for localizing
 faulty processes in message passing programs, aiming at reducing
 developers' efforts. MPI-PD helps us to identify the source of
 failure from a number of observed errors by automatically checking
 communication errors during program execution. If MPI-PD observes any
 communication errors, it then generates a trace file, backtraces
 communication dependencies and points out potentially faulty
 processes in the event graph visualization.

 MPI-PD reduces the amount of debugging information before visualizing
 and investigating it by using post-mortem performance debuggers and
 source-level debuggers, respectively. 
 Therefore, we think that appropriate combined use of these tools
 is a good choice for debugging message passing programs.

 \section*{Acknowledgements}
 This work was partly supported by JSPS Grant-in-Aid for Young
 Researchers (B)(15700030), for Scientific
 Research (C)(2)(14580374), JSPS Research for the Future Program
 JSPS-RFTF99I00903, and Network Development Laboratories, NEC.
 We are also grateful to the anonymous reviewers
 for their valuable comments.

\bibliographystyle{plain}
\bibliography{main}

\end{document}